\documentclass[aps,prb,showpacs,twocolumn,amsmath,amssymb,superscriptaddress,letterpaper]{revtex4}
\usepackage{times}
\usepackage{amsfonts}
\usepackage{mathrsfs}
\usepackage{graphicx}
\usepackage{dcolumn}
\usepackage{bm}
\usepackage{color}

\usepackage[colorlinks,bookmarks=false,citecolor=blue,linkcolor=red,urlcolor=blue]{hyperref}
\def\be{\begin{equation}}       \def\ee{\end{equation}}
\def\bea{\begin{eqnarray}}      \def\eea{\end{eqnarray}}

\begin{document}
\title{$\beta$-CuI: a Dirac semimetal without surface Fermi arcs}

\author{Congcong Le}
\affiliation{Kavli Institute of Theoretical Sciences, and CAS Center for Excellence in Topological Quantum Computation, University of Chinese Academy of Sciences, Beijing, 100190, China}
\affiliation{Beijing National Laboratory for Condensed Matter Physics, Institute of Physics, Chinese Academy of Sciences, Beijing 100190, China}

\author{Xianxin Wu}\email{xianxinwu@gmail.com}
\affiliation{Institute for Theoretical Physics and Astrophysics,
Julius-Maximilians University of W\"urzburg, Am Hubland, D-97074 W\"urzburg, Germany}

\author{Shengshan Qin}
\affiliation{Beijing National Laboratory for Condensed Matter Physics, Institute of Physics, Chinese Academy of Sciences, Beijing 100190, China}
\affiliation{University of Chinese Academy of Science, Beijing 100049, China}

\author{Yinxiang Li}
\affiliation{Beijing National Laboratory for Condensed Matter Physics, Institute of Physics, Chinese Academy of Sciences, Beijing 100190, China}

\author{Ronny Thomale}
\affiliation{Institute for Theoretical Physics and Astrophysics,
Julius-Maximilians University of W\"urzburg, Am Hubland, D-97074 W\"urzburg, Germany}

\author{Fuchun Zhang}
\affiliation{Kavli Institute of Theoretical Sciences, and CAS Center for Excellence in Topological Quantum Computation, University of Chinese Academy of Sciences, Beijing, 100190, China}

\author{Jiangping Hu  }\email{jphu@iphy.ac.cn} \affiliation{Beijing National Laboratory for Condensed Matter Physics, Institute of Physics, Chinese Academy of Sciences, Beijing 100190, China}
\affiliation{Kavli Institute of Theoretical Sciences, and CAS Center for Excellence in Topological Quantum Computation, University of Chinese Academy of Sciences, Beijing, 100190, China}\affiliation{Collaborative Innovation Center of Quantum Matter, Beijing, 100049,China}

\date{\today}

\begin{abstract}
Anomalous surface states with Fermi arcs are commonly considered to be a fingerprint of Dirac semimetals (DSMs).  In contrast to Weyl semimetals, however, Fermi arcs of DSMs are not topologically protected. Using first-principles calculations, we predict that $\beta$-CuI is a peculiar DSM whose surface states form closed Fermi pockets instead of Fermi arcs. In such a fermiological Dirac semimetal, the deformation mechanism from Fermi arcs to Fermi pockets stems from a large cubic term preserving all crystal symmetries, and the small energy difference between the surface and bulk Dirac points. The cubic term in $\beta$-CuI, usually  negligible in prototypical DSMs,  becomes relevant because of the particular crystal structure. As such, we establish a concrete material example manifesting the lack of topological protection for surface Fermi arcs in DSMs.

\end{abstract}

\pacs{75.85.+t, 75.10.Hk, 71.70.Ej, 71.15.Mb}

\maketitle

Topological semimetals including Dirac semimetals (DSMs), Weyl semimetals (WSMs), and nodal line semimetals have been attracting enormous attention in contemporary research\cite{Chiu2016,Armitage2017}, exhibiting a plethora of exotic phenomena\cite{xu2011,Wan2011,Balents2011,Wang2012,Wang2013, Parameswaran2014,Huang2015,LiangT2015,BaumY2015,XiangJ2015,WengHM2015W,HuangSM2015}.  In particular, the surface states of such semimetals commonly feature open Fermi arcs  rather than closed Fermi pockets.  The principal existence of Fermi arcs appears robust against potential bulk band hybridizations, and have been confirmed by theoretical calculations as well as experimental observations  in  all  type-\uppercase\expandafter{\romannumeral1} and type-\uppercase\expandafter{\romannumeral2}\cite{Soluyanov2015} WSM and DSM materials studied so far\cite{ChangTR2017,HuangH2017,Yan2017,NohHJ2017,FeiFC2017,Le2017}.

The topological protection of non-degenerate surface Fermi arcs in WSMs traces back to topological invariants enforcing the connection between Berry flux monopoles of opposite charge, which is realized by pairs of bulk Weyl cones projected to a given surface.   In view of DSMs, however, it has been pointed out recently~\cite{Kargariana2016,Lu2017} that the doubly degenerate Fermi arcs on side surfaces are {\it not} topologically protected, and that a cubic term preserving all crystal symmetries can deform Fermi arcs into closed Fermi surfaces, yielding a state we call fermiological DSM. In all DSMs (Na$_3$Bi and Cd$_3$As$_2$) known so far, such a cubic term is  negligible, so that doubly degenerate Fermi arcs always appear at the surfaces.

In this Letter, we predict that $\beta$-CuI\cite{Shan2009} is the first proposed instance of a fermiological Dirac semimetal, exhibiting closed Fermi surfaces instead of Fermi arcs on its side surfaces. The band inversion, which can be greatly enhanced with compressive strain along $c$ axis, happens between the bonding states of Cu-$4s$ orbitals and I-$5p_{x,y}$ orbitals. It generates both three-dimensional (3D) topological semimetal and 3D topological insulator phases. A crystal symmetry preserving cubic term, which is usually expected to be negligible in previous DSM materials, is found to be considerably large because of the unique atomic arrangements in $\beta$-CuI, in sharp contrast to conventional Dirac semimetals such as Na$_3$Bi and Cd$_3$As$_2$.
In particular,  the small energy difference between surface  and bulk Dirac points causes  a flat surface state   along the $\Gamma$-Z direction.
In this flat surface state, the cubic term  can introduce a gap  for $k_z\neq 0$ to deform Fermi arcs into a closed Fermi surface.  Our study provides a concrete material example to illustrate the lack of topological protection of surface Fermi arcs in DSMs. The corresponding consequences in ARPES and quantum oscillation measurements are also discussed.

\begin{figure}
\centerline{\includegraphics[width=0.4\textwidth]{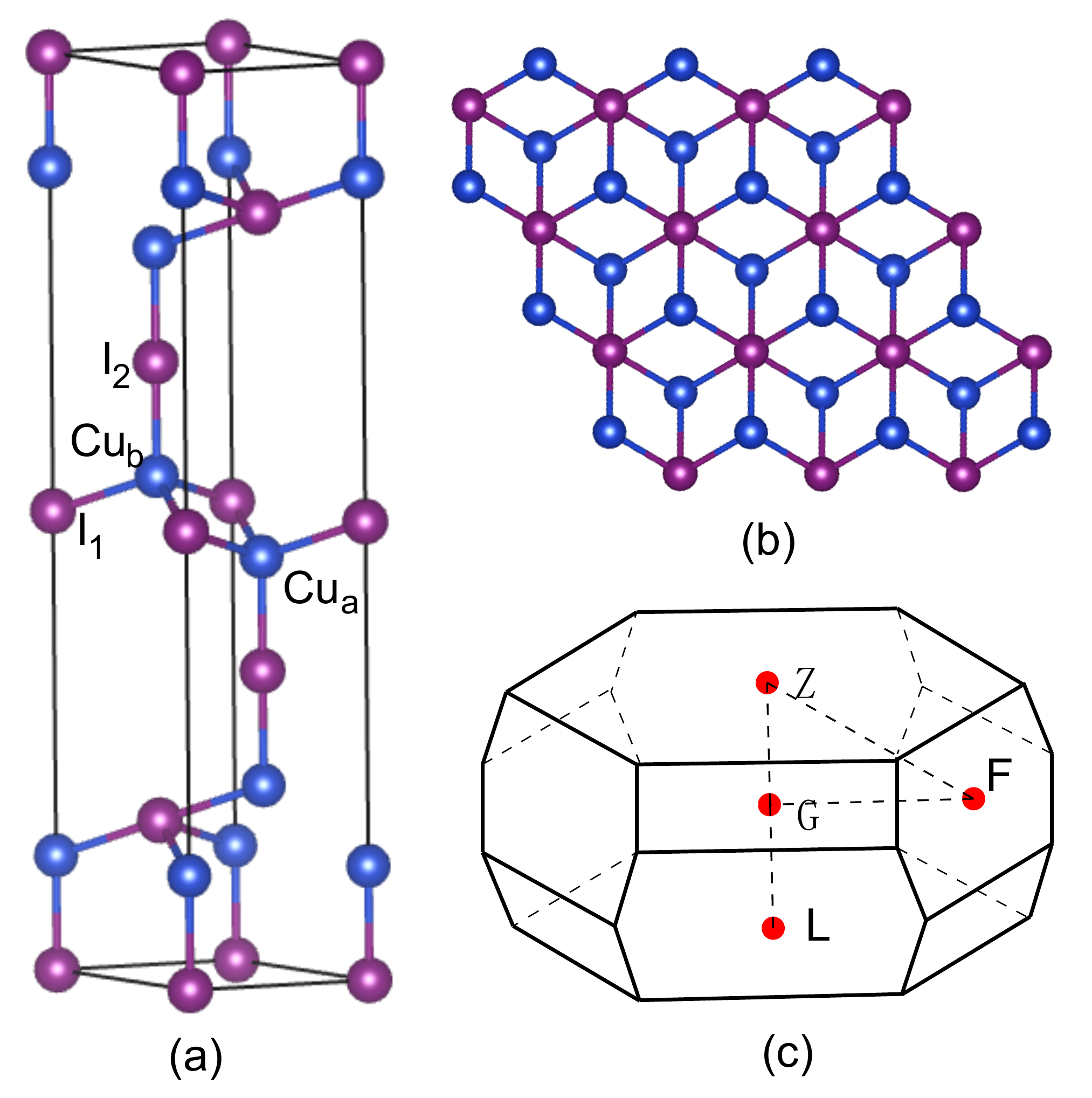}}
\caption{{\bf Crystal structure and primitive Brillouin zone for $\beta$-CuI.} (a) Crystal structure of $\beta$-CuI. Cu-I$_1$-Cu trilayers stacking along c axis are connected by I$_2$ ions. (b) Top view of Crystal structure of Cu-I$_1$-Cu trilayers. I$_1$ is octahedrally coordinated by six Cu atoms, which generate a negative crystal field. (c) Primitive Brillouin zone for $\beta$-CuI.
\label{structurecui} }
\end{figure}

{\it \bf{Crystal Structure}}
The crystal chemistry of cuprous iodide (CuI) is characterized by three stable structural phases $\alpha$, $\beta$ and $\gamma$ \cite{Shan2009}. Here, we focus on the topologically nontrivial properties of the $\beta$ phase. The crystal structure of $\beta$-CuI with the space group $R\bar{3}m$  is shown in Fig.\ref{structurecui}(a)\cite{Shan2009}. According to the chemical environment, the I ions can be classified as I$_1$ and I$_2$. I$_1$ is octahedrally coordinated by six Cu atoms, and I$_2$ is coordinated by only two Cu ions parallel to the $c$ axis, resulting in a strong negative crystal field for the I$_1$ p orbitals and I$_2$ p$_z$ orbital. As shown in Fig.\ref{structurecui}(a), the Cu-I$_1$-Cu form trilayer structures, and are are connected by I$_2$ ions along the $c$ axis. In the following calculations, we adopt the experimental structural parameters in Ref.\onlinecite{Shan2009}.

\begin{figure}
\centerline{\includegraphics[width=0.45\textwidth]{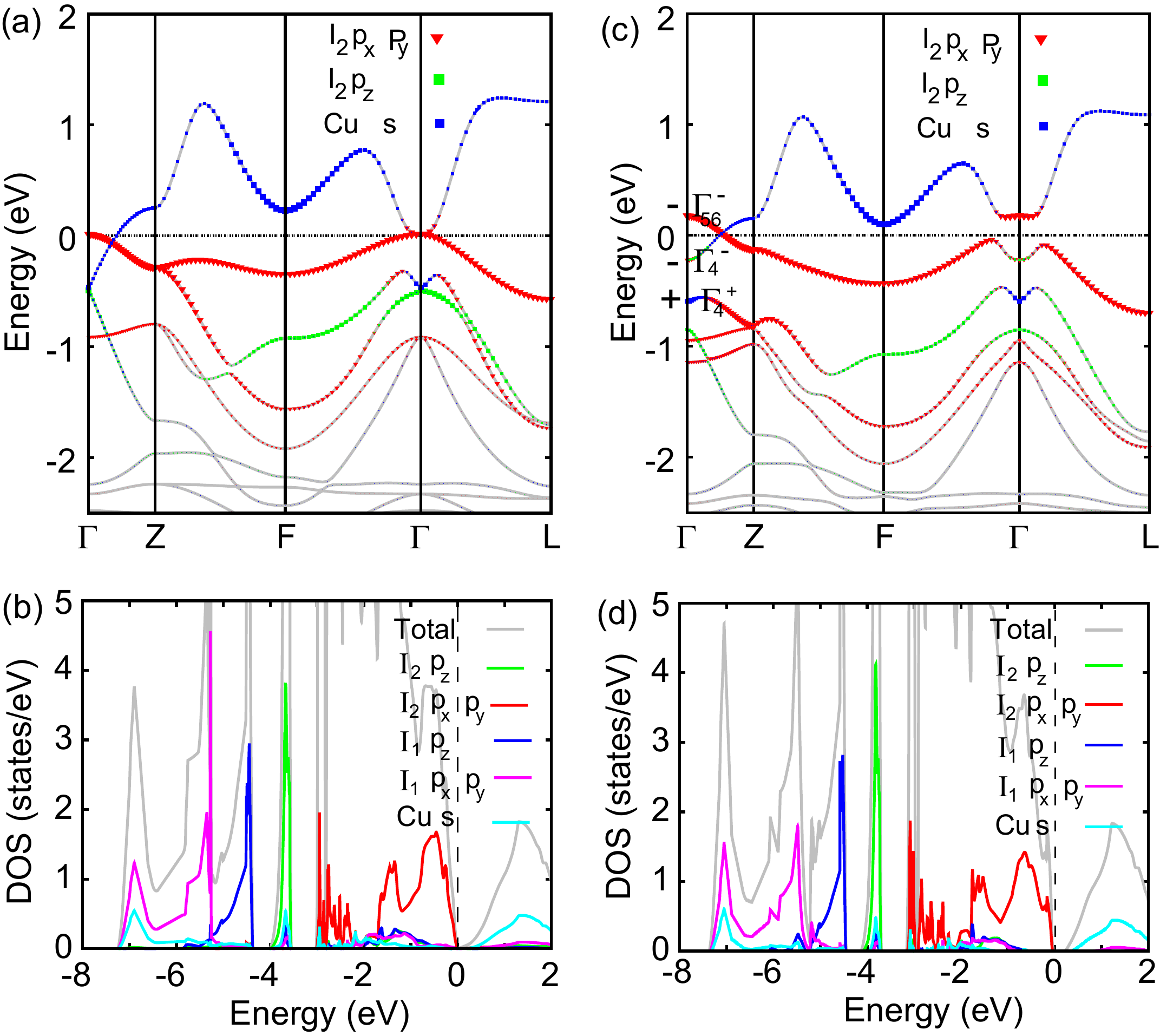}}
\caption{{\bf Band structures and Density of states for $\beta$-CuI without SOC and with SOC.} (a) and (b) Band structures and density-of-states of $\beta$-CuI without SOC. The band inversion happens between I$_2$-$5p_{x/y}$ and Cu-$4s$ orbitals, and Dirac points are located in $\Gamma$Z near the Fermi level. (c) and (d) Band structures and density-of-states of $\beta$-CuI with SOC. The I-5p$_{x,y}$ bands in $\Gamma Z$ line are splitted, and the band inversion at $\Gamma$ point is further enhanced to be 0.77 eV. The orbital weights are represented by the areas of circles and triangles. The parities of the eigenstates and the irreducible representations of bands at the $\Gamma$ point near the Fermi level are shown.
\label{banddos} }
\end{figure}

{\it \bf{ Electronic structure }}
The band structure and density of states (DOS) for $\beta$-CuI are displayed in Fig.\ref{banddos}. Due to the monovalence of Cu, the $d$ orbitals of Cu are fully filled and located at about -2.5 eV. The $p$ orbitals of I$_1$ and the $p_z$ orbital of I$_2$ lie far below the Fermi level because of the strongly negative crystal field. Near the Fermi level, the valence and conduction bands are predominantly attributed to the I$_2$-$5p_{x/y}$ and Cu-$4s$ orbitals. The most prominent feature in the band structure is that at the $\Gamma$ point, the Cu-$4s$ band is lower than the I-5p$_{x,y}$ bands by about 0.47 eV, and that there is a Dirac cone along $\Gamma Z$ line near the Fermi level, as shown in Fig.\ref{banddos}(a). Due to strong spin orbital coupling (SOC) in I ions, we further take SOC into consideration in our calculations. As shown in Fig.\ref{banddos}(c), the I-5p$_{x,y}$ bands in $\Gamma Z$ line split into $\Gamma_{56}$ ($|j_z=\pm\frac{3}{2}\rangle$) and $\Gamma_{4}$  ($|j_z=\pm\frac{1}{2}\rangle$) bands, and the band inversion at $\Gamma$ point is further enhanced to 0.77 eV. As the generalized gradient approximation occasionally tends to underestimate band gaps, we further assert the avenue of band inversion by hybrid functional HSE calculations, and also find that the gap can be greatly enhanced through compressive strain along the $c$ axis (see supplementary material). Furthermore, as the two crossing bands along $\Gamma Z$ line belong to different irreducible representations as distinguished by $C_3$ rotational symmetry around the $z$ axis, this indicates that the 3D Dirac cones near the Fermi level are stable. Notably, the Cu-$4s$ and I-5p$_{x,y}$ $|j_z=\pm\frac{1}{2}\rangle$ bands have the same $\Gamma_{4}$  irreducible representation, which leads to a full gap opening around -0.4 eV. As the parities of Cu-$4s$ and I-$5p$ bands are opposite at the $\Gamma$ and $Z$ point, band inversion will drive the system into a topologically nontrivial phase. Due to the presence of three-dimensional inversion symmetry in $\beta$-CuI, we can calculate $Z_2$ topological invariants by analyzing the parity eigenstates at high symmetry points\cite{Fu2007}. The parity of the eigenstates near the Fermi level at $\Gamma$ and $Z$ points are displayed in Fig.\ref{banddos}(c). According to our calculations, CuI is a topologically nontrivial semimetal, with 3D $Z_2$ invariants given by $(1;000)$. Furthermore, setting the chemical potential to -0.4 eV, the system is located in a topological insulator phase with nontrivial $Z_2$ invariants. In total, we thus find that band inversion generates both topological semimetal and topological insulator phases.

\begin{figure}
\centerline{\includegraphics[width=0.5\textwidth]{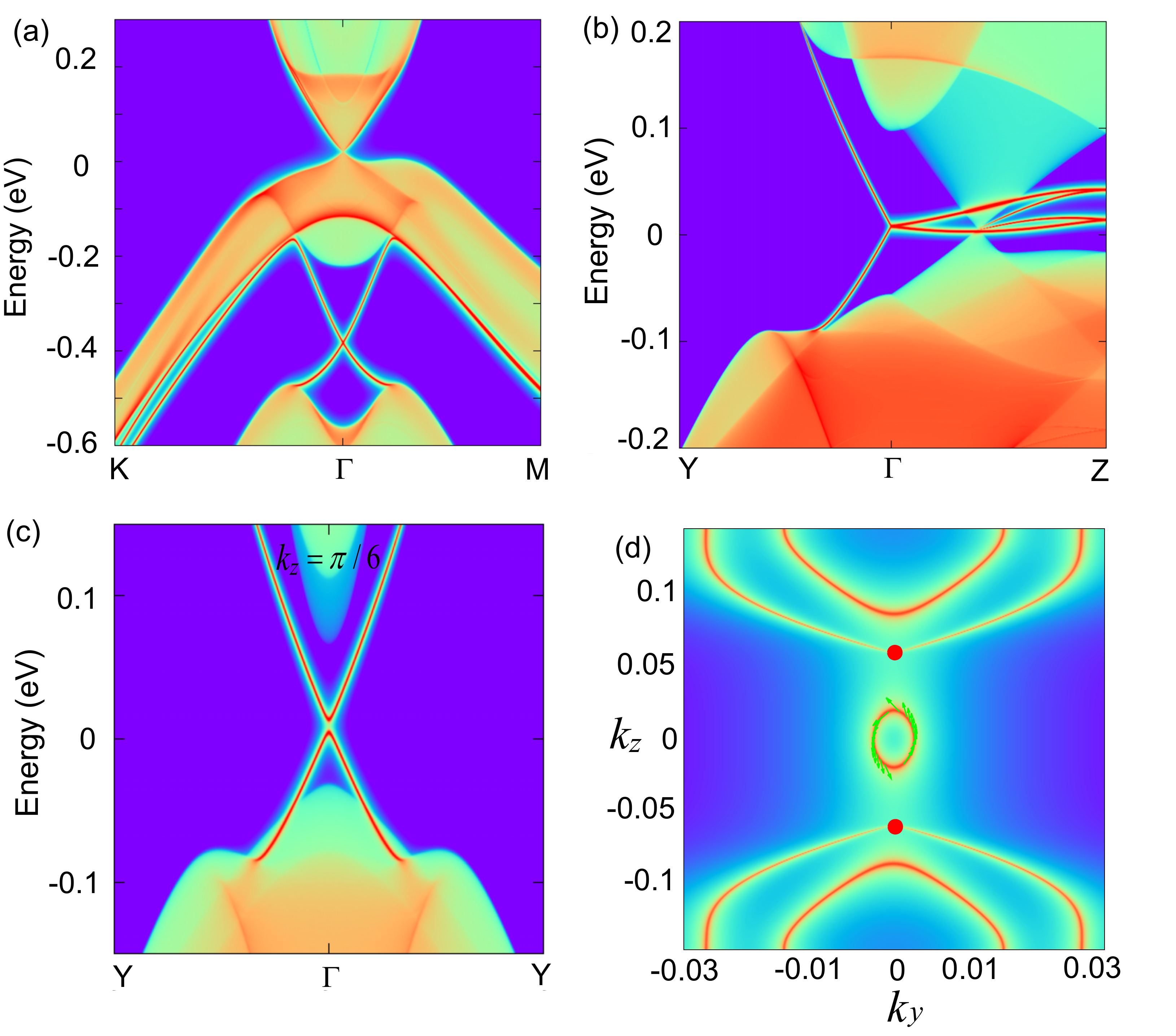}}
\caption{{\bf (001) and (100) surface states and Fermi surfaces for $\beta$-CuI.}  (a) and (b) Projected surface states of $\beta$-CuI for (001) and (100) surfaces in the conventional cell. On (001) surface a surface Dirac cone exist around -0.4 eV and on (100) surface the two surface states exhibit non-monotonic dispersion along $\Gamma Z$ and vanish at the projection of bulk Dirac points. (c) projected surface states of $\beta$-CuI for (100) surface at $k_z=\frac{\pi}{6}$ plane, where surface states are gapped. (d) Fermi surface at the energy of bulk Dirac points for (100) surface. One closed nontrivial Fermi pocket with spin helical texture (shown by green arrows) is centered around $k_z=0$. The closed pocket does not pass through the projections of the bulk Dirac points (red circles), illustrating that Fermi arcs are absent.
\label{edge} }
\end{figure}

Because of bulk-edge correspondence, a topologically nontrivial bulk state is accompanied by gapless surface states. For CuI, those can be obtained by calculating the surface Green function of the semi-infinite system through an iterative procedure\cite{Sancho1984,Sancho1985}. Fig.\ref{edge} (a) shows the edge states on the (001) surface. Interestingly, a surface Dirac cone exist around -0.4 eV stemming from the nontrivial topological insulator phase, and the corresponding Fermi surface is a closed circle with a left-handed spin texture (see supplementary material). The surface states of the (100) surface in the conventional cell are shown in Fig.\ref{edge} (b). The energy difference $\Delta$ between the surface Dirac point at $\Gamma$ and the projections of the bulk Dirac points is extremely small, yielding flat surface states along $\Gamma Z$, in sharp contrast to conventional DSMs. Despite the band folding along the $\Gamma Z$ direction, we find that the two surface states vanish at the projection of bulk Dirac points and exhibit non-monotonic dispersion along $\Gamma Z$. Furthermore, the lower surface state first sinks below the energy level $E_{D}$ of the bulk Dirac points, then raises above it, and finally bends down to saturate at it, resulting in three crossing points for $k_y=0$ at $E_D$. The corresponding Fermi surface of (100) surface at $E_D$ is shown in Fig.\ref{edge}(d). There is one closed nontrivial Fermi pocket centered around $k_z=0$ and two trivial pockets around $k_z=\pi$, which originates from the nontrivial $Z_2$ invariant in the $k_z=0$ plane and the trivial $Z_2$ invariant in the $k_z=\pi$ plane, respectively. The closed Fermi pocket around $k_z=0$ does not pass through the projections of the bulk Dirac points (denoted by red circles),  illustrating that Fermi arcs are absent. Furthermore, the surface states at an exemplary amount of $k_z=\pi/6$ lower than the location $k_{zD}$ of the Dirac point, which exhibit gap opening, are shown in Fig.\ref{edge} (c). We find that the obtained surface states are gapped for all $k_z$ except $k_z=0$, which crucially contributes to deforming Fermi arcs into a closed Fermi surface.

{\it \bf{ Effective Hamiltonian }}
To characterize the low-energy effective Hamiltonian around the $\Gamma$ point, which is helpful to understand the origin of the surface Fermi arc breakdown, we adopt the perspective of theory of invariants\cite{Liu2010}. From the band structure, the states around $\Gamma$ are mainly attributed to I$_2$-$5p_{x,y}$ and Cu-$4s$ orbitals, and thus these orbitals can be used to construct the basis. Further considering the inversion symmetry in the system, it is convenient to combine these orbitals to form the eigenstates of the inversion symmetry, which are given by
\begin{eqnarray}
 |P^{\pm}_{\alpha}\rangle&=&\frac{1}{\sqrt{2}}(|\text{I}_{\alpha}\rangle\pm|\text{I}^{\prime}_{\alpha}\rangle),\nonumber
 \\
 |S^{\pm}\rangle&=&\frac{1}{\sqrt{2}}(|\text{Cu}_{s}\rangle\pm|\text{Cu}^{\prime}_{s}\rangle),
\end{eqnarray}
where the superscript denotes the parity, $\alpha=p_{x,y}$, and the I (Cu) as well as I$^{\prime}$ (Cu$^{\prime}$) atoms are related by inversion symmetry. We focus on the low-energy states near the bulk Dirac point. After further taking into account SOC in the  atomic picture, we can choose $|S^{+}, \frac{1}{2}\rangle$, $|P^{-} ,\frac{3}{2}\rangle$, $|S^{+} ,-\frac{1}{2}\rangle$, $|P^{-} ,-\frac{3}{2}\rangle$ as the basis in $\bm{k}\cdot\bm{p}$ theory to construct the effective Hamiltonian around the $\Gamma$ point. The Hamiltonian to third order in $\bm{k}$ reads
\begin{eqnarray}
H_{eff}(\mathbf{k})  &=& H_0+H_1+H_2\nonumber\\
 H_0&=&\epsilon(\mathbf{k})+ M(\mathbf{k})\sigma_0\tau_3-A(\mathbf{k}_{\parallel})(k_x \sigma_3\tau_2+k_y \sigma_0\tau_1) \nonumber\\
 H_1&=& (D_2+D_3k^2_z)(-k_x \sigma_1\tau_2+k_y \sigma_2\tau_2) \nonumber\\
 H_2&=& -D_1k_z[(k^2_x-k^2_y) \sigma_1\tau_2+2k_xk_y \sigma_2\tau_2],
\label{eq2ds}
\end{eqnarray}
where the Pauli matrices $\bm{\sigma}$ act in spin and $\bm{\tau}$ in orbital space, $k_{\pm}=k_x{\pm}ik_y$, $\epsilon_{\mathbf{k}}=C_0+C_1k^2_{z}+C_{2}(k^2_{x}+k^2_{y})$, $M(\mathbf{k})=M_{0}-M_{1}k^2_{z}-M_{2}(k^2_{x}+k^2_{y})$, $A(\mathbf{k}_{\parallel})=A_{0}+A_{1}k^2_z$, $D(\mathbf{k})=ik_{+}(D_2+D_3k^2_z)$, and $\tilde{D}(\mathbf{k})=iD_1 k_z k^2_{-}$. The anti-diagonal terms contain first-order and third-order terms, which have often been omitted in previous studies, but turn out to be of great importance in $\beta$-CuI$_2$. The energy dispersion of the Hamiltonian for the DSM is $E(k)=\epsilon_{\mathbf{k}}\pm\sqrt{M(\mathbf{k})^2+A^2k_{+}k_{-}+|D(\mathbf{k})+\tilde{D}(\mathbf{k})|^2}$, resulting in two band crossing points (0,0, $\pm k_{zD}$) along $\Gamma$-Z line with $k_{zD}=\sqrt{\frac{M_0}{M_1}}$.  By fitting the bands of the effective model with those of DFT calculation around the $\Gamma$ point, the parameters in the effective model are given by $C_0=-0.2070$ eV, $C_1=2.0445$ eV$\cdot$\AA$^2$~, $C_2=12.8481$ eV$\cdot$\AA$^2$~, $M_0=-0.3855 $ eV, $M_1=-6.8288$ eV$\cdot$\AA$^2$~, $M_2=-37.4544$ eV$\cdot$\AA$^2$, $A_0=4.0035$ eV$\cdot$\AA~, $A_1=-1629.0242$ eV$\cdot$\AA$^2$~, $D_1=167.799$ eV$\cdot$\AA$^3$~, $D_2=2.8549$ eV$\cdot$\AA~, and $D_3=-1668.6306$ eV$\cdot$\AA$^3$~. In $\beta$-CuI, we find that the coefficients in the anti-diagonal terms are considerably large, and thus cannot be omitted.

\begin{figure}
\centerline{\includegraphics[width=0.5\textwidth]{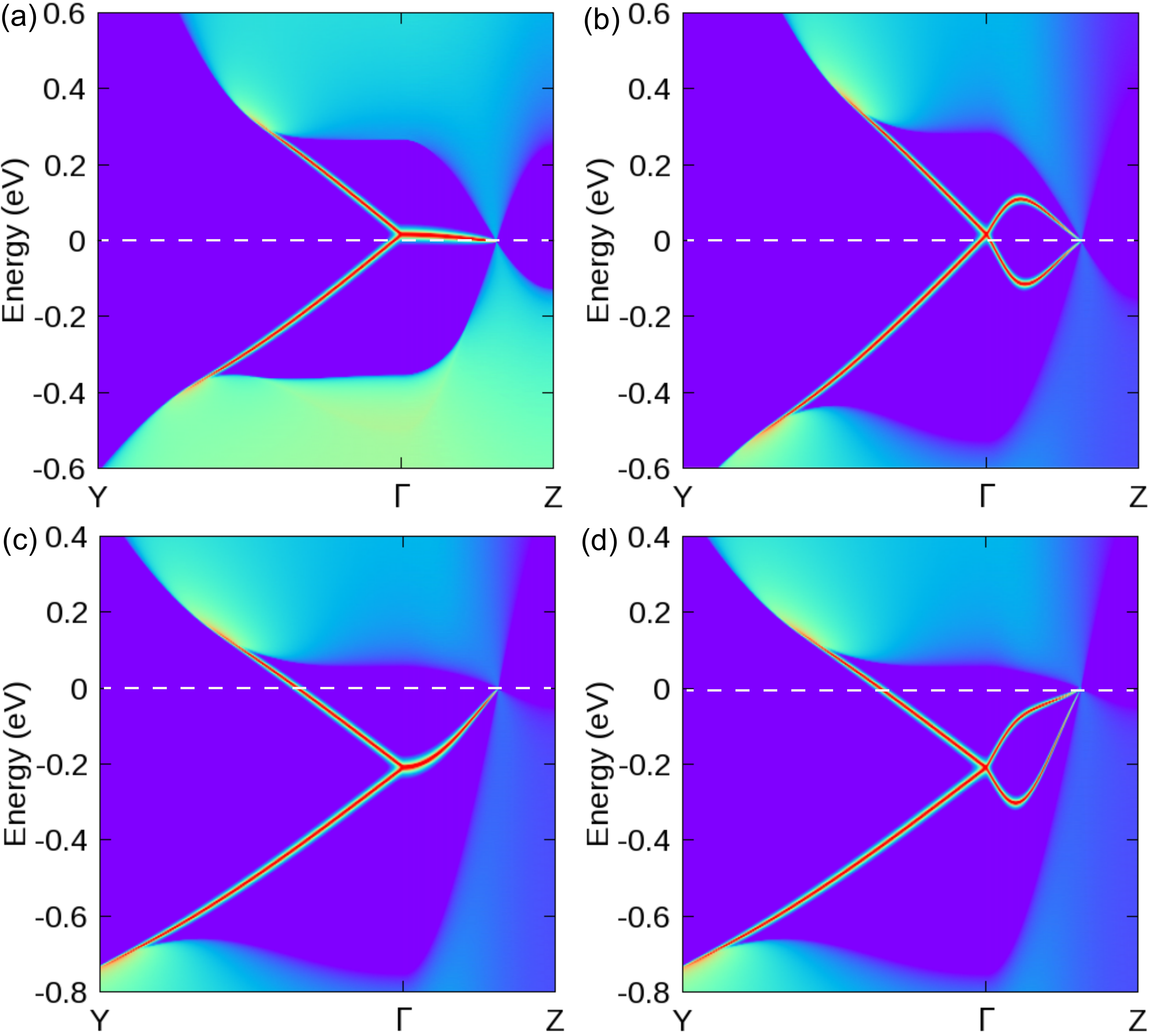}}
\caption{{\bf  (100) surface states from the effective Hamiltonian with different $D_1$ and $\Delta$.} Calculated (100) surfaces states from $H_0+H_1$ (a) and $H_0+H_1+H_2$ (b) with a small $\Delta$. Taking $H_2$ into consideration, the Fermi arcs deform into a closed Fermi pocket. Calculated surface states from $H_0+H_1$ (c) and $H_0+H_1+H_2$ (d) with a large $\Delta$. The effect of cubic term is weakened due to the large $\Delta$ and Fermi arcs can still exist. Both large cubic term and small $\Delta$ are crucial to the absence of side Fermi arcs in Dirac semmimetal.
\label{kpband} }
\end{figure}

In Ref.\onlinecite{Kargariana2016}, the Fermi arcs on (100) surface have been shown to be not protected by symmetry and can in principle be absent. Still, the effective Hamiltonian $H_0$, up to second order in $\bf{k}$, can give robust surface Fermi arcs. Therefore, $H_0$ must have additional symmetries, which are to some degree artificial and not enforeced for DSM materials. Consider a pseudo time reversal symmetry $\mathcal{T}$ in 2D, which can be defined as $\mathcal{T}=i\sigma_2\tau_3\cdot K$. Under this operation, the Hamiltonian for $H(k_x,k_y,k_{z_0})$ at a fixed $k_{z_0}$ plane transforms as $\mathcal{T}H(k_x,k_y,k_{z_0})\mathcal{T}^{-1}=H(-k_x,-k_y,k_{z_0})$. It can be easily verified that the Hamiltonian $H_0$ and $H_1$ are invariant under the operation $\mathcal{T}$. This symmetry, not preserved for the generic realistic system but only for the Hamiltonian $H_0$ and its side surfaces, can protect gapless surface states for any $k_z<k_{z_D}$ planes. The energy difference between the surface Dirac point and bulk Dirac point is given by
 \begin{eqnarray}
\Delta=(\frac{C_2}{M_2}-\frac{C_1}{M_1})M_0.
 \end{eqnarray}
 The corresponding prototypical surface states on (100) surface along Y-$\Gamma$-Z for $H_0+H_1$, with a small $\Delta$, are shown in Fig.\ref{kpband}(a), where the energy of two degenerate flat surface states decreases monotonically with increasing momentum along $\Gamma Z$. As a consequence of the latter, there are only two points in $k_y=0$ on Fermi surface at $E_D$, that is, the projected bulk Dirac points, and Fermi arcs can robustly appear on the (100) surface. It is, however, the cubic $H_2$ term that breaks this artificial symmetry, and naturally introduces gap openings for any $k_z$ except $k_z=0$ where fundamental time reversal symmetry is kept. Taking $H_2$ into consideration, two surface states split, as shown in Fig.\ref{kpband}(b), and the prominent feature is that both surface states exhibit a non-monotonic band dispersion along $\Gamma$-Z, which generates additional two points in $k_y=0$ at $E_D$ in the surface state. As such, the Fermi arcs deform into a closed Fermi pocket (see supplementary material), bearing some similarity to a 3D topological insulator surface state. Further increasing coefficients in $H_1$, we find that this will reduce the splitting of surface states along $\Gamma$-Z. Adding either inversion symmetry or time reversal symmetry breaking, DSMs become WSMs, and Fermi arcs are known to be robust (see supplementary material). In the presence of both these symmetries, however, the cubic term explicates how Fermi arcs on the surface of a DSM are not topologically protected.

We now turn to a detailed analysis why despite the above finding, the hallmark of DSM materials discovered previously, such as Na$_3$Bi and Cd$_3$As$_2$, still has been the appearance of seemingly robust Fermi arcs. We conjecture that this is attributed to a small coefficient of the cubic term in $H_2$ along with a small $\Delta$. How does this change for $\beta$-CuI~? We start by analysing the origin of the $H_2$ term, which corresponds to the coupling of $|P^{-}, \frac{3}{2}\rangle$ and $|S^{+}, -\frac{1}{2}\rangle$. The only process generating this coupling in $\beta$-CuI can be summarized as
\begin{eqnarray}
|p^{I_2}_{x/y,\sigma}\rangle \stackrel{\lambda_I}{\longrightarrow}  |p^{I_2}_{z,\bar{\sigma}}\rangle  \stackrel{t_1}{\longrightarrow}  |s^{Cu_b}_{\bar{\sigma}}\rangle \stackrel{t_2}{\longrightarrow}  |s^{Cu_a}_{\bar{\sigma}}\rangle,
\end{eqnarray}
where $\sigma=\uparrow,\downarrow$ labels the spin. The hybridization process is as follows: first, $|p^{I_2}_{x/y,\sigma}\rangle$ couples strongly with $|p^{I_2}_{z,\bar{\sigma}}\rangle$ due to strong atomic SOC in I atoms; as there is a strong $\sigma$ bond between I$_2$ and Cu$_b$ , $|p^{I_2}_{z,\bar{\sigma}}\rangle$ can strongly hybridize with $|s^{Cu_b}_{\bar{\sigma}}\rangle$; because of the short distance between Cu$_a$ and Cu$_b$, $|s^{Cu_a}_{\bar{\sigma}}\rangle$ and $|s^{Cu_b}_{\bar{\sigma}}\rangle$ exhibit considerable coupling; finally, the $|P^{-}, \frac{3}{2}\rangle$ can couple indirectly via $|S^{+}, -\frac{1}{2}\rangle$, and the coupling constant $D_1$ is proportional to $\lambda_I t_1t_2$. In $\beta$-CuI, all three parameters are large, and hence they generate a considerable $D_1$. While the microscopic mechanism explained above is derived for a specific material, it potentially applies to a series of other DSMs, certainly for the ones originating from band inversion between $|j_z=\pm \frac{3}{2}\rangle$ and $|j_z=\pm \frac{1}{2}\rangle$ states with opposite parities. In Na$_3$Bi and Cd$_3$As$_2$, even though SOC is even stronger than for CuI, the second and third step of the process are considerably weakened because of the weak bonding between cations and anions, indicating that the $D_1$ parameter is small there. In addition, $\Delta$ in Na$_3$Bi is much larger than that of $\beta$-CuI, which also weakens the effect of the cubic term. Fig.\ref{kpband}(c) shows the surface states with a large $\Delta$, where surface states exhibit a large dispersion along $\Gamma$-Z. Including the same cubic term as in Fig.\ref{kpband}(b), the band splitting weakens, and Fermi arcs can still exist in this case. Thus, small $\Delta$ is another prerequisite to impose the absence of Fermi arcs. The $H_1$ term in $\beta$-CuI preserves $\mathcal{T}$ symmetry, and, if dominant, can substantially suppress the gap opening for the surface states. The first term in $D(\bm{k})$, however, only involves inplane coupling, and is weak because of no immediate microscopic foundation in real space; the second term includes $k^2_z$, and as such, in comparison to $H_2$ has much weaker effect for small $k_z$. Therefore, the combined appearance of large $D_1$ as well as small $\Delta$ in $\beta$-CuI triggers the breakdown of surface Fermi arcs.

{\it \bf{ Discussion}}
We elaborate on experimental evidences deriving from the breakdown of surface Fermi arcs due to the significant cubic term. Firstly, aiming at the effect of the cubic term in the bulk, the inplane energy dispersion for a specific $k_z$ is $E(k)=\epsilon_{\mathbf{k}}\pm\sqrt{f_1+f_2|k|^2+f_3|k|^4}$. If $D_1$ is large, the coefficient $f_3=M^2_2+D^2_1k^2_z$ should exhibit noticeable $k_z$ dependence, which could be identified upon fitting the band structure against ARPES measurements. Secondly, as the splitting of (100) surface states along $\Gamma$-Z is directly related to the cubic term, this splitting can likewise be obtained in ARPES, and is expected to be relatively large as well as strongly $k_z$-dependent. In addition, the change of nature of the surface states from arcs to closed Fermi pockets hints at immediate experimental implications.  Firstly, terminating Fermi arcs and closed Fermi surfaces exhibit qualitative differentiable shape differences in ARPES measurements. For the former, when two Fermi arcs meet at the projection of bulk Dirac points, there is a singular change in slope, whereas for the latter, the closed Fermi surface has a smooth curvature everywhere and does not pass through the projections of bulk Dirac points. Secondly, the distinct behavior of surface Fermi arcs versus closed surface Fermi pockets in quantum oscillation measurements can be used to contrast them. In the former case, the quantum oscillation frequency $F_s$ is strongly dependent on the sample thickness due to the Weyl orbits\cite{Potter2014,Mol2014}. In triangle-shaped samples, quantum oscillations can be even unobservable in experiment\cite{Mol2014}. In the latter case, fermions acquire a measurable Berry phase of $\pi$ as they encircle the Fermi contour, similar to topological insulators. In contrast to the former case, quantum oscillations can exist in triangle-shaped samples\cite{Ren2010,Qu2010,Analytis2010} and exhibit weak thickness dependence.

{\it \bf{Conclusion}}
 Based on first-principles calculations we predict that $\beta$-CuI is the first topological unconventional Dirac semimetal exhibiting closed Fermi surfaces instead of Fermi arcs on its side surfaces. The theoretical discovery of $\beta$-CuI provide an explicit proof that the Fermi arcs in DS are not topologically protected.  Our study also  suggests that halide compounds can be a fertile ground to explore novel topological properties.

{\it \bf{Method} }

Our calculations are performed using density functional theory (DFT) as implemented in the Vienna ab initio simulation package (VASP) code \cite{Kresse1993,Kresse1996,Kresse1996B}. The Perdew-Burke-Ernzerhof (PBE) exchange-correlation functional and the projector-augmented-wave (PAW) approach are used. Throughout the work, the cutoff energy is set to be 500 eV for expanding the wave functions into plane-wave basis. In the calculation, the Brillouin zone is sampled in the k space within Monkhorst-Pack scheme\cite{MonkhorstPack}. On the basis of the equilibrium structure, the k mesh used is $6\times6\times6$ and $10\times10\times2$ for primitive and conventional cell, respectively.

{\it \bf{ References} }
\bibliographystyle{nature}

\begin{thebibliography}{10}
\expandafter\ifx\csname url\endcsname\relax
  \def\url#1{\texttt{#1}}\fi
\expandafter\ifx\csname urlprefix\endcsname\relax\def\urlprefix{URL }\fi
\providecommand{\bibinfo}[2]{#2}
\providecommand{\eprint}[2][]{\url{#2}}

\bibitem{Chiu2016}
\bibinfo{author}{Chiu, C.-K.}, \bibinfo{author}{Teo, J. C.~Y.},
  \bibinfo{author}{Schnyder, A.~P.} \& \bibinfo{author}{Ryu, S.}
\newblock \bibinfo{title}{Classification of topological quantum matter with
  symmetries}.
\newblock \textit{\bibinfo{journal}{Rev. Mod. Phys.}}
  \textbf{\bibinfo{volume}{88}}, \bibinfo{pages}{035005}
  (\bibinfo{year}{2016}).
\newblock
  \urlprefix\url{https://link.aps.org/doi/10.1103/RevModPhys.88.035005}.

\bibitem{Armitage2017}
\bibinfo{author}{Armitage, N.~P.}, \bibinfo{author}{Mele, E.~J.} \&
  \bibinfo{author}{Ashvin~Vishwanath, A.}
\newblock \bibinfo{title}{Weyl and dirac semimetals in three dimensional
  solids}.
\newblock \textit{\bibinfo{journal}{arXiv:}} \bibinfo{pages}{1705.01111}
  (\bibinfo{year}{2017}).

\bibitem{xu2011}
\bibinfo{author}{Xu, G.}, \bibinfo{author}{Weng, H.}, \bibinfo{author}{Wang,
  Z.}, \bibinfo{author}{Dai, X.} \& \bibinfo{author}{Fang, Z.}
\newblock \bibinfo{title}{Chern semimetal and the quantized anomalous hall
  effect in ${\mathrm{hgcr}}_{2}{\mathrm{se}}_{4}$}.
\newblock \textit{\bibinfo{journal}{Phys. Rev. Lett.}}
  \textbf{\bibinfo{volume}{107}}, \bibinfo{pages}{186806}
  (\bibinfo{year}{2011}).
\newblock
  \urlprefix\url{https://link.aps.org/doi/10.1103/PhysRevLett.107.186806}.

\bibitem{Wan2011}
\bibinfo{author}{Wan, X.}, \bibinfo{author}{Turner, A.~M.},
  \bibinfo{author}{Vishwanath, A.} \& \bibinfo{author}{Savrasov, S.~Y.}
\newblock \bibinfo{title}{Topological semimetal and fermi-arc surface states in
  the electronic structure of pyrochlore iridates}.
\newblock \textit{\bibinfo{journal}{Phys. Rev. B}}
  \textbf{\bibinfo{volume}{83}}, \bibinfo{pages}{205101}
  (\bibinfo{year}{2011}).
\newblock \urlprefix\url{https://link.aps.org/doi/10.1103/PhysRevB.83.205101}.

\bibitem{Balents2011}
\bibinfo{author}{Balents, L.}
\newblock \bibinfo{title}{Weyl electrons kiss}.
\newblock \textit{\bibinfo{journal}{Physics}} \textbf{\bibinfo{volume}{4}},
  \bibinfo{pages}{36} (\bibinfo{year}{2011}).

\bibitem{Wang2012}
\bibinfo{author}{Wang, Z.} \textit{et~al.}
\newblock \bibinfo{title}{Dirac semimetal and topological phase transitions in
  $\mathrm{A}_{3}\mathrm{Bi}$
  ($\mathrm{A}=\mathrm{Na},\mathrm{K},\mathrm{Rb}$)}.
\newblock \textit{\bibinfo{journal}{Phys. Rev. B}}
  \textbf{\bibinfo{volume}{85}}, \bibinfo{pages}{195320}
  (\bibinfo{year}{2012}).
\newblock \urlprefix\url{https://link.aps.org/doi/10.1103/PhysRevB.85.195320}.

\bibitem{Wang2013}
\bibinfo{author}{Wang, Z.}, \bibinfo{author}{Weng, H.}, \bibinfo{author}{Wu,
  Q.}, \bibinfo{author}{Dai, X.} \& \bibinfo{author}{Fang, Z.}
\newblock \bibinfo{title}{Three-dimensional dirac semimetal and quantum
  transport in $\mathrm{Cd}_{3}\mathrm{As}_{2}$}.
\newblock \textit{\bibinfo{journal}{Phys. Rev. B}}
  \textbf{\bibinfo{volume}{88}}, \bibinfo{pages}{125427}
  (\bibinfo{year}{2013}).
\newblock \urlprefix\url{https://link.aps.org/doi/10.1103/PhysRevB.88.125427}.

\bibitem{Parameswaran2014}
\bibinfo{author}{Parameswaran, S.~A.}, \bibinfo{author}{Grover, T.},
  \bibinfo{author}{Abanin, D.~A.}, \bibinfo{author}{Pesin, D.~A.} \&
  \bibinfo{author}{Vishwanath, A.}
\newblock \bibinfo{title}{Probing the chiral anomaly with nonlocal transport in
  three-dimensional topological semimetals}.
\newblock \textit{\bibinfo{journal}{Phys. Rev. X}}
  \textbf{\bibinfo{volume}{4}}, \bibinfo{pages}{031035} (\bibinfo{year}{2014}).
\newblock \urlprefix\url{https://link.aps.org/doi/10.1103/PhysRevX.4.031035}.

\bibitem{Huang2015}
\bibinfo{author}{Huang, X.} \textit{et~al.}
\newblock \bibinfo{title}{Observation of the chiral-anomaly-induced negative
  magnetoresistance in 3d weyl semimetal $\mathrm{TaAs}$}.
\newblock \textit{\bibinfo{journal}{Phys. Rev. X}}
  \textbf{\bibinfo{volume}{5}}, \bibinfo{pages}{031023} (\bibinfo{year}{2015}).
\newblock \urlprefix\url{https://link.aps.org/doi/10.1103/PhysRevX.5.031023}.

\bibitem{LiangT2015}
\bibinfo{author}{Liang, T.} \textit{et~al.}
\newblock \bibinfo{title}{Ultrahigh mobility and giant magnetoresistance in the
  dirac semimetal $\mathrm{Cd}_{3}\mathrm{As}_{2}$}.
\newblock \textit{\bibinfo{journal}{Nat. Mater.}}
  \textbf{\bibinfo{volume}{14}}, \bibinfo{pages}{280} (\bibinfo{year}{2014}).
\newblock \urlprefix\url{http://dx.doi.org/10.1038/nmat4143}.

\bibitem{BaumY2015}
\bibinfo{author}{Baum, Y.}, \bibinfo{author}{Berg, E.},
  \bibinfo{author}{Parameswaran, S.~A.} \& \bibinfo{author}{Stern, A.}
\newblock \bibinfo{title}{Current at a distance and resonant transparency in
  weyl semimetals}.
\newblock \textit{\bibinfo{journal}{Phys. Rev. X}}
  \textbf{\bibinfo{volume}{5}}, \bibinfo{pages}{041046} (\bibinfo{year}{2015}).
\newblock \urlprefix\url{https://link.aps.org/doi/10.1103/PhysRevX.5.041046}.

\bibitem{XiangJ2015}
\bibinfo{author}{Xiong, J.} \textit{et~al.}
\newblock \bibinfo{title}{Evidence for the chiral anomaly in the dirac
  semimetal $\mathrm{Na}_{3}\mathrm{Bi}$}.
\newblock \textit{\bibinfo{journal}{Science}} \textbf{\bibinfo{volume}{350}},
  \bibinfo{pages}{413--416} (\bibinfo{year}{2015}).
\newblock
  \urlprefix\url{http://science.sciencemag.org/content/sci/350/6259/413.full.pdf}.

\bibitem{WengHM2015W}
\bibinfo{author}{Weng, H.}, \bibinfo{author}{Fang, C.}, \bibinfo{author}{Fang,
  Z.}, \bibinfo{author}{Bernevig, B.~A.} \& \bibinfo{author}{Dai, X.}
\newblock \bibinfo{title}{Weyl semimetal phase in noncentrosymmetric
  transition-metal monophosphides}.
\newblock \textit{\bibinfo{journal}{Phys. Rev. X}}
  \textbf{\bibinfo{volume}{5}}, \bibinfo{pages}{011029} (\bibinfo{year}{2015}).
\newblock \urlprefix\url{https://link.aps.org/doi/10.1103/PhysRevX.5.011029}.

\bibitem{HuangSM2015}
\bibinfo{author}{Huang, S.-M.} \textit{et~al.}
\newblock \bibinfo{title}{A weyl fermion semimetal with surface fermi arcs in
  the transition metal monopnictide $\mathrm{Ta}\mathrm{As}$ class}.
\newblock \textit{\bibinfo{journal}{Nat Commun.}} \textbf{\bibinfo{volume}{6}},
  \bibinfo{pages}{7373} (\bibinfo{year}{2015}).
\newblock \urlprefix\url{http://dx.doi.org/10.1038/ncomms8373}.

\bibitem{Soluyanov2015}
\bibinfo{author}{Soluyanov, A.~A.} \textit{et~al.}
\newblock \bibinfo{title}{Type-ii weyl semimetals}.
\newblock \textit{\bibinfo{journal}{Nature}} \textbf{\bibinfo{volume}{527}},
  \bibinfo{pages}{495} (\bibinfo{year}{2015}).
\newblock \urlprefix\url{http://dx.doi.org/10.1038/nature15768}.

\bibitem{ChangTR2017}
\bibinfo{author}{Chang, T.-R.} \textit{et~al.}
\newblock \bibinfo{title}{Type-ii symmetry-protected topological dirac
  semimetals}.
\newblock \textit{\bibinfo{journal}{Phys. Rev. Lett.}}
  \textbf{\bibinfo{volume}{119}}, \bibinfo{pages}{026404}
  (\bibinfo{year}{2017}).
\newblock
  \urlprefix\url{https://link.aps.org/doi/10.1103/PhysRevLett.119.026404}.

\bibitem{HuangH2017}
\bibinfo{author}{Huang, H.}, \bibinfo{author}{Zhou, S.} \&
  \bibinfo{author}{Duan, W.}
\newblock \bibinfo{title}{Type-ii dirac fermions in the $\mathrm{PtSe}_{2}$
  class of transition metal dichalcogenides}.
\newblock \textit{\bibinfo{journal}{Phys. Rev. B}}
  \textbf{\bibinfo{volume}{94}}, \bibinfo{pages}{121117}
  (\bibinfo{year}{2016}).
\newblock \urlprefix\url{https://link.aps.org/doi/10.1103/PhysRevB.94.121117}.

\bibitem{Yan2017}
\bibinfo{author}{Yan, M.} \textit{et~al.}
\newblock \bibinfo{title}{Lorentz-violating type-ii dirac fermions in
  transition metal dichalcogenide $\mathrm{Pt}\mathrm{Te}_{2}$}.
\newblock \textit{\bibinfo{journal}{Nat Commun.}} \textbf{\bibinfo{volume}{8}},
  \bibinfo{pages}{257} (\bibinfo{year}{2017}).
\newblock \urlprefix\url{https://doi.org/10.1038/s41467-017-00280-6}.

\bibitem{NohHJ2017}
\bibinfo{author}{Noh, H.-J.} \textit{et~al.}
\newblock \bibinfo{title}{Experimental realization of type-ii dirac fermions in
  a ${\mathrm{pdte}}_{2}$ superconductor}.
\newblock \textit{\bibinfo{journal}{Phys. Rev. Lett.}}
  \textbf{\bibinfo{volume}{119}}, \bibinfo{pages}{016401}
  (\bibinfo{year}{2017}).
\newblock
  \urlprefix\url{https://link.aps.org/doi/10.1103/PhysRevLett.119.016401}.

\bibitem{FeiFC2017}
\bibinfo{author}{Fei, F.} \textit{et~al.}
\newblock \bibinfo{title}{Nontrivial berry phase and type-ii dirac transport in
  the layered material $\mathrm{PdTe}_{2}$}.
\newblock \textit{\bibinfo{journal}{Phys. Rev. B}}
  \textbf{\bibinfo{volume}{96}}, \bibinfo{pages}{041201}
  (\bibinfo{year}{2017}).
\newblock \urlprefix\url{https://link.aps.org/doi/10.1103/PhysRevB.96.041201}.

\bibitem{Le2017}
\bibinfo{author}{Le, C.} \textit{et~al.}
\newblock \bibinfo{title}{Three-dimensional topological critical dirac
  semimetal in $\mathrm{AMgBi}$
  ($\mathrm{A}=\mathrm{K},\mathrm{Rb},\mathrm{Cs}$}.
\newblock \textit{\bibinfo{journal}{Phys. Rev. B}}
  \textbf{\bibinfo{volume}{96}}, \bibinfo{pages}{115121}
  (\bibinfo{year}{2017}).
\newblock \urlprefix\url{https://link.aps.org/doi/10.1103/PhysRevB.96.115121}.

\bibitem{Kargariana2016}
\bibinfo{author}{Kargarian, M.}, \bibinfo{author}{Randeria, M.} \&
  \bibinfo{author}{Lu, Y.-M.}
\newblock \bibinfo{title}{Are the surface fermi arcs in dirac semimetals
  topologically protected?}
\newblock \textit{\bibinfo{journal}{Proc. Natl. Acad. Sci.}}
  \bibinfo{pages}{201524787} (\bibinfo{year}{2016}).

\bibitem{Lu2017}
\bibinfo{author}{Kargarian, M.}, \bibinfo{author}{Lu, Y.-M.} \&
  \bibinfo{author}{Randeria, M.}
\newblock \bibinfo{title}{Deformation and stability of surface states in dirac
  semimetals}.
\newblock \textit{\bibinfo{journal}{arXiv:}} \bibinfo{pages}{1712.03982}
  (\bibinfo{year}{2017}).

\bibitem{Shan2009}
\bibinfo{author}{Shan, Y.} \textit{et~al.}
\newblock \bibinfo{title}{Description of the phase transitions of cuprous
  iodide}.
\newblock \textit{\bibinfo{journal}{J. Alloys Compd.}}
  \textbf{\bibinfo{volume}{477}}, \bibinfo{pages}{403--406}
  (\bibinfo{year}{2009}).

\bibitem{Fu2007}
\bibinfo{author}{Fu, L.} \& \bibinfo{author}{Kane, C.~L.}
\newblock \bibinfo{title}{Topological insulators with inversion symmetry}.
\newblock \textit{\bibinfo{journal}{Phys. Rev. B}}
  \textbf{\bibinfo{volume}{76}}, \bibinfo{pages}{045302}
  (\bibinfo{year}{2007}).
\newblock \urlprefix\url{https://link.aps.org/doi/10.1103/PhysRevB.76.045302}.

\bibitem{Sancho1984}
\bibinfo{author}{Sancho, M.~L.}, \bibinfo{author}{Sancho, J.~L.} \&
  \bibinfo{author}{Rubio, J.}
\newblock \bibinfo{title}{Quick iterative scheme for the calculation of
  transfer matrices: application to $\mathrm{Mo}$ (100)}.
\newblock \textit{\bibinfo{journal}{J. Phys. F}} \textbf{\bibinfo{volume}{14}},
  \bibinfo{pages}{1205} (\bibinfo{year}{1984}).

\bibitem{Sancho1985}
\bibinfo{author}{Sancho, M.~L.}, \bibinfo{author}{Sancho, J.~L.},
  \bibinfo{author}{Sancho, J.~L.} \& \bibinfo{author}{Rubio, J.}
\newblock \bibinfo{title}{Highly convergent schemes for the calculation of bulk
  and surface green functions}.
\newblock \textit{\bibinfo{journal}{J. Phys. F}} \textbf{\bibinfo{volume}{15}},
  \bibinfo{pages}{851} (\bibinfo{year}{1985}).

\bibitem{Liu2010}
\bibinfo{author}{Liu, C.-X.} \textit{et~al.}
\newblock \bibinfo{title}{Model hamiltonian for topological insulators}.
\newblock \textit{\bibinfo{journal}{Phys. Rev. B}}
  \textbf{\bibinfo{volume}{82}}, \bibinfo{pages}{045122}
  (\bibinfo{year}{2010}).
\newblock \urlprefix\url{https://link.aps.org/doi/10.1103/PhysRevB.82.045122}.

\bibitem{Potter2014}
\bibinfo{author}{Potter, A.~C.}, \bibinfo{author}{Kimchi, I.} \&
  \bibinfo{author}{Vishwanath, A.}
\newblock \bibinfo{title}{Quantum oscillations from surface fermi arcs in weyl
  and dirac semimetals}.
\newblock \textit{\bibinfo{journal}{Nat Commun.}} \textbf{\bibinfo{volume}{5}},
  \bibinfo{pages}{5161} (\bibinfo{year}{2014}).
\newblock \urlprefix\url{http://dx.doi.org/10.1038/ncomms6161}.

\bibitem{Mol2014}
\bibinfo{author}{Moll, P. J.~W.} \textit{et~al.}
\newblock \bibinfo{title}{Transport evidence for fermi-arc-mediated chirality
  transfer in the dirac semimetal $\mathrm{Cd}_{3}\mathrm{As}_{2}$}.
\newblock \textit{\bibinfo{journal}{Nature}} \textbf{\bibinfo{volume}{535}},
  \bibinfo{pages}{266} (\bibinfo{year}{2016}).
\newblock \urlprefix\url{http://dx.doi.org/10.1038/nature18276}.

\bibitem{Ren2010}
\bibinfo{author}{Ren, Z.}, \bibinfo{author}{Taskin, A.~A.},
  \bibinfo{author}{Sasaki, S.}, \bibinfo{author}{Segawa, K.} \&
  \bibinfo{author}{Ando, Y.}
\newblock \bibinfo{title}{Large bulk resistivity and surface quantum
  oscillations in the topological insulator
  ${\text{bi}}_{2}{\text{te}}_{2}\text{Se}$}.
\newblock \textit{\bibinfo{journal}{Phys. Rev. B}}
  \textbf{\bibinfo{volume}{82}}, \bibinfo{pages}{241306}
  (\bibinfo{year}{2010}).
\newblock \urlprefix\url{https://link.aps.org/doi/10.1103/PhysRevB.82.241306}.

\bibitem{Qu2010}
\bibinfo{author}{Qu, D.-X.}, \bibinfo{author}{Hor, Y.~S.},
  \bibinfo{author}{Xiong, J.}, \bibinfo{author}{Cava, R.~J.} \&
  \bibinfo{author}{Ong, N.~P.}
\newblock \bibinfo{title}{Quantum oscillations and hall anomaly of surface
  states in the topological insulator $\mathrm{Bi}_{2}\mathrm{Te}_{3}$}.
\newblock \textit{\bibinfo{journal}{Science}} \textbf{\bibinfo{volume}{329}},
  \bibinfo{pages}{821--824} (\bibinfo{year}{2010}).
\newblock
  \urlprefix\url{http://science.sciencemag.org/content/sci/329/5993/821.full.pdf}.

\bibitem{Analytis2010}
\bibinfo{author}{Analytis, J.~G.} \textit{et~al.}
\newblock \bibinfo{title}{Two-dimensional surface state in the quantum limit of
  a topological insulator}.
\newblock \textit{\bibinfo{journal}{Nat. Phys.}} \textbf{\bibinfo{volume}{6}},
  \bibinfo{pages}{960} (\bibinfo{year}{2010}).
\newblock \urlprefix\url{http://dx.doi.org/10.1038/nphys1861}.

\bibitem{Kresse1993}
\bibinfo{author}{Kresse, G.} \& \bibinfo{author}{Hafner, J.}
\newblock \bibinfo{title}{Ab initio molecular dynamics for liquid metals}.
\newblock \textit{\bibinfo{journal}{Phys. Rev. B}}
  \textbf{\bibinfo{volume}{47}}, \bibinfo{pages}{558--561}
  (\bibinfo{year}{1993}).
\newblock \urlprefix\url{https://link.aps.org/doi/10.1103/PhysRevB.47.558}.

\bibitem{Kresse1996}
\bibinfo{author}{Kresse, G.} \& \bibinfo{author}{Furthmüller, J.}
\newblock \bibinfo{title}{Efficiency of ab-initio total energy calculations for
  metals and semiconductors using a plane-wave basis set}.
\newblock \textit{\bibinfo{journal}{Comput. Mater. Sci.}}
  \textbf{\bibinfo{volume}{6}}, \bibinfo{pages}{15--50} (\bibinfo{year}{1996}).

\bibitem{Kresse1996B}
\bibinfo{author}{Kresse, G.} \& \bibinfo{author}{Furthmüller, J.}
\newblock \bibinfo{title}{Efficient iterative schemes for ab initio
  total-energy calculations using a plane-wave basis set}.
\newblock \textit{\bibinfo{journal}{Phys. Rev. B}}
  \textbf{\bibinfo{volume}{54}}, \bibinfo{pages}{11169--11186}
  (\bibinfo{year}{1996}).
\newblock \urlprefix\url{https://link.aps.org/doi/10.1103/PhysRevB.54.11169}.

\bibitem{MonkhorstPack}
\bibinfo{author}{Monkhorst, H.~J.} \& \bibinfo{author}{Pack, J.~D.}
\newblock \bibinfo{title}{Special points for brillouin-zone integrations}.
\newblock \textit{\bibinfo{journal}{Phys. Rev. B}}
  \textbf{\bibinfo{volume}{13}}, \bibinfo{pages}{5188--5192}
  (\bibinfo{year}{1976}).
\newblock \urlprefix\url{https://link.aps.org/doi/10.1103/PhysRevB.13.5188}.

\end{thebibliography}

{\it \bf{Acknowledgments}}

We thank Chen Fang and Lunhui Hu for helpful discussion. This work is supported by the Ministry of Science and Technology
of China 973 program (Grants No. 2015CB921300
and No. 2017YFA0303100), National Science Foundation of
China (Grant No. NSFC-11334012), and the Strategic Priority
Research Program of CAS (Grant No. XD-B07000000). This work is supported in part by the Key Research Program of the Chinese Academy of Sciences (Grant No. XDPB08-4) and  NSFC grant: 11674278. The work in W\"urzburg is supported by ERC-StG-TOPOLECTRICS-336012, DFG-SFB 1170, and DFG-SPP 1666.

{\it \bf{Author contributions}}

C.-C.L. initiated the project. C.-C.L. and X.-X.W. performed the \emph{ab-initio} calculations. All the authors participate in discussion and writing of the manuscript. X.-X.W. and J.-P.H. supervised the project.

{\it \bf{Competing interests}}
The authors declare no competing financial interests

\end{document}